\documentstyle[11pt,newpasp,twoside,epsf]{article}
\def\edcomment#1{\iffalse\marginpar{\raggedright\sl#1\/}\else\relax\fi}
\reversemarginpar

\newcommand\der{{\rm d}} 		
\newcommand{\msun}{{\rm M_{\odot}}}

\begin{document}
\title{A New Evolutionary Picture for CVs and LMXBs\\
  II. The Impact of Thermal-Timescale Mass Transfer}

\author{K. Schenker and A.~R. King} 

\affil{Theoretical Astrophysics Group, University of Leicester, 
  Leicester, \mbox{LE1 7RH}, U.K.}


\begin{abstract}
Depending on the outcome of pre-CV formation, mass transfer
may set in under thermally unstable conditions in a significant number
of systems. Full computations have shown that such an early phase of
thermal-timescale mass transfer usually leads to ordinary looking CVs,
but these do also show some unusual properties (e.g.\ chemical
anomalies in later stages).

Rather than investigating the common envelope evolution leading to
pre-CVs, we study the properties of multiple evolutionary tracks
starting with a phase of thermal-timescale mass transfer.
Apart from fitting unusual CVs (like AE Aqr), global properties of the
CV population as a whole give indications that this is indeed the
channel where many CVs come from.
\end{abstract}


\section{Introduction}

The standard picture of cataclysmic variable (CV) evolution has in
spite of its many benefits serious problems, some of which have been
discussed in the previous contribution. If we adopt the new picture
sketched therein, it is worth to investigate in detail whether some of 
these problems have indeed disappeared (without generating an equal
number of new ones :-). 

As an additional starting point, we take a closer look at some of the
individual, strange objects among known CVs. First of all there is 
{AE Aqr}, a rapidly spinning intermediate polar at 
$P_{\rm orb} \simeq 10 \, {\rm hr}$, which apparently has no accretion
disk at all. Another puzzling system is {V1309 Ori}, a polar at
$P_{\rm orb} \simeq 8 \, {\rm hr}$. Finally we would like to 
mention {V485 Cen}, a system that harbours a probably He-rich 
donor (Augustijn et al.\ 1996) at $P_{\rm orb} \simeq 1 \, {\rm hr}$ and
may be considered as an example of a binary in-between CVs and AM CVn
systems. 
We will see that all of these can be understood as part of the group of
non-standard CVs emerging from higher mass donors, which can
be much more evolved (but still on the main sequence) when mass
transfer commences.

The important question asked in this contribution would therefore be: 
How do more massive donor stars (progenitors) in a CV evolve in the
light of less effective common envelope (CE) evolution? Can we make at
least some of all these loose ends in CV evolution come together?


\section{Features of Thermal-Timescale Mass Transfer Evolution}

\begin{figure}
\plotfiddle{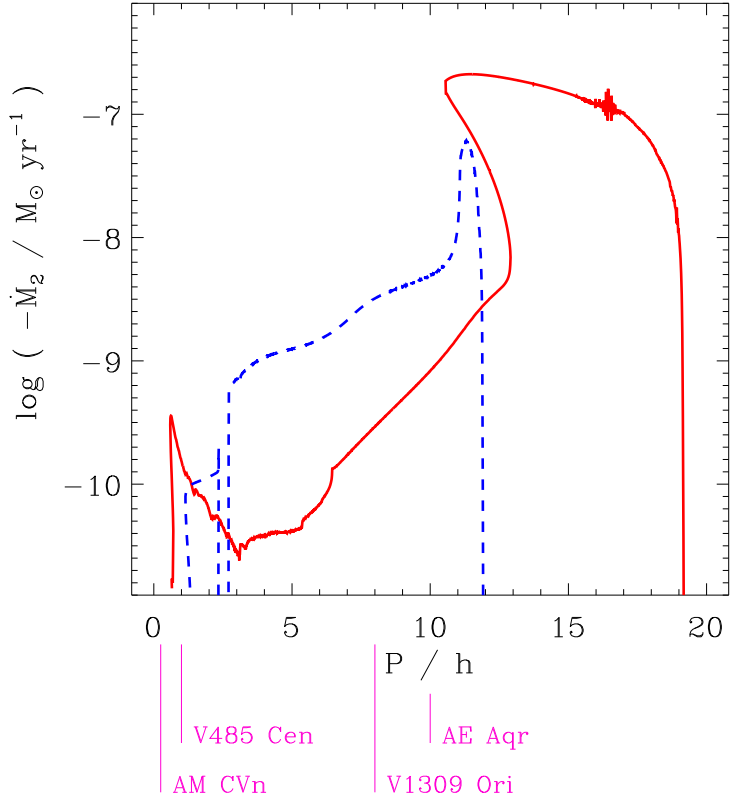}{10cm}{0}{70}{70}{-175}{20}
\caption{Evolution of mass transfer rate over orbital period for two
different cases of thermal-timescale mass transfer: Track 1 
(broken line) for a weak case shows an almost normal behaviour after an 
initial phase, while the extreme case track 2 (full line, with lower
primary mass and a much more evolved donor of the same mass) shows a
typical S-shaped curve during the TTMT, and no period gap but a lower mass
transfer rate all the way down to an orbital period minimum below $1
\, {\rm hr}$. 
Additional labels mark the orbital period of various interesting
systems discussed in the text.}
\end{figure}

Of fundamental importance is the question of thermal stability. 
When applying standard concepts and assumptions (cf.\ Ritter 1996) this 
there is an upper limit $q_{\rm crit}$ on the mass ratio $M_2 / M_1$
for main sequence (MS) donors, above which mass is transferred on
roughly $M_2$'s thermal timescale, i.e.\ leads to Thermal-Timescale
Mass Transfer (TTMT).
Such phases of TTMT can be computed with a stellar evolution code
(Cyg X-2 (Kolb et al.\ 2000), SN Ia study (Langer et al.\ 2000),
and LMXBs (Podsiadlowski et al.\ 2001). 

An additional second upper limit is due to delayed dynamical
instability (DDI, Webbink 1977), thus systems with
$q_{\rm crit} < q < q_{\rm DDI}$ would undergo TTMT and 
may become ordinary CVs afterwards (or {\em extraordinary} ones, like
AE Aqr and similar cases).\footnote{Our current unability to follow
the dynamical evolution in any reliable way forces us to exclude all
cases of dynamically unstable mass transfer, although their actual
role and importance in close binary evolution is far from being
settled.} 

This leaves a range of initial donor star masses where instead of the
normal, stable CV evolution we get a more complex one including an
initial, thermally unstable phase of mass transfer. 
In general, both the sensitivity of such a phase on details of the
mass loss, and the wider range of possible nuclear evolution (due to
the shorter MS life-time for more massive stars) lead to a
much more complex individual mass transfer history. Let us take a
closer look with the help of two very different examples.


\subsection{Two exemplary evolution tracks}

\begin{figure}
\plotfiddle{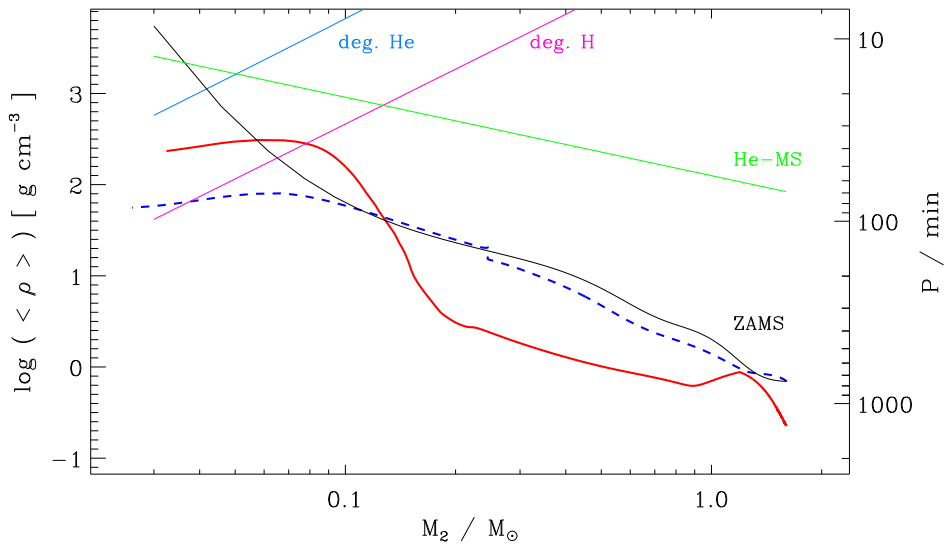}{7cm}{90}{55}{55}{220}{-40}
\caption{Mean density (and orbital period) of the same two tracks as
in Fig.~1 are shown in comparison with various theoretical lines as
labelled. 
Note that while track 1 (broken line) is reasonably well represented
by a ZAMS donor until the very late phase around the bounce period,
track 2 (full line) is extremely different: It appears to be much more
evolved (lower mean density), until the He-core gets exposed, leading
to a more compact star filling its Roche lobe down to ultra-short
orbital periods.}
\end{figure}

The dashed curve in Fig.~1 shows a case of `weak' TTMT. The
initial donor star mass $M_{2,i} = 1.6 \, \msun$, filling its Roche
lobe in a $12 \, {\rm hr}$ binary with a $1.4 \, \msun$ primary star
(the sequence actually shows a low-mass X-ray binary with a neutron
star accretor). The relatively short orbital period indicates the low
age of the donor, which still has around $56 \%$ hydrogen left in the
core at this moment. In contrast, the thick full curve gives the
other end of the possible scale of TTMT evolution, showing a secondary
of the same mass ($1.6 \, \msun$), but much further evolved: here the
central hydrogen is down to $\sim 5 \%$, which explains the much
larger radius and hence the initial orbital period of $19 \, {\rm hr}$.
This binary has a $0.7 \, \msun$ white dwarf primary, so the initial
mass ratio is quite large leading to a very violent initial phase of
TTMT, whereas the other example merely shown an initial `hump' of
enhanced $\dot{M}_2$, hardly modifying the subsequent evolution when
compared to say a similar track starting with a $1.3 \, \msun$ donor. 

One of the common assumptions regarding CVs is, that the donors are
essentially unevolved main sequence stars, i.e.\ their mass-radius
relation should be close to theoretical ZAMS models. Figure 2
illustrates how different our second track behaves in this respect,
showing the mean density $\left< \rho \right>$ evolution over donor
mass for the two examples. Other lines in the plot show the ZAMS, a
fit curve for the location of the He-MS, plus two lines derived from
the assumption of completely degenerate stars consisting of either
solar (H-rich) material or pure He.
A common expression 
\begin{equation}
    \left< \rho \right> \simeq 
	115 \, P_{\rm hr}^{-2} \, {\rm g} {\rm cm}^{-3}
    \, ,
\end{equation}    
which is based on Paczy\'nski's approximation for the Roche radius,
allows to translate $\left< \rho \right>$ into orbital period (e.g.\ King 1988).

It is quite obvious from these lines, that while the dashed curve of
track 1 nicely follows the ZAMS line, a similar assumption for the
radii -- say using $M_2$-$P_{\rm orb}$ relation -- of the other
sequence would give completely wrong results. 
From high to low masses, the donor is
very oversized (because it was near the end of its MS life), forced to
shrink by the initially fast mass transfer (TTMT), then relaxing back
towards its larger equilibrium radius. Around $0.2 \, \msun$ the old
core becomes exposed, and the star is rapidly becoming more compact
than a normal MS mass of similar mass (heading from below the ZAMS
towards the He-MS, roughly), until degeneracy becomes important and
the star turns towards the degenerate He line (at a much shorter
orbital period than the corresponding H-rich star, which turns towards
the degenerate H-line).


\subsection{Pre-CV evolution}

The diagrams shown in Figs.~3 \& 4 of the previous contribution (King
\& Schenker 2002) are illustrating a scenario when there are two kinds
of systems emerging from the common envelope evolution forming the WD
in a future CV.
In the first `normal' group of systems, some form of
magnetic braking is operating on relatively low-mass secondaries to
bring them into contact by angular momentum losses. These will become
ordinary CVs, hosting effectively unevolved donor stars (due to their
low masses and the short timescale associated with a sufficiently 
effective braking).
The other group contains more massive secondaries, that may have no
additional braking mechanism at all, but start filling their Roche
lobe simply due to expansion {\em on the main sequence}.
Detailed properties of such a bimodal population depend severely on
binary formation, CE evolution, and magnetic braking (or any other
angular momentum loss that might be operating). As long as huge
uncertainties in the combination of these effects persist, we should
try and reverse the problem by trying to make predictions from
observed features of the pre-CV and CV population for the required 
initial states when mass transfer began in each of the systems. 

In addition to the discussion in King \& Schenker (2002),
we would like to make the point here why our track 2 is to be expected
to occur assuming a post-CE situation vaguely similar to the
simplistic assumption going into the pre-CV plot shown before: A
post-CE orbital period of around $20 .. 30 \, {\rm hr}$ allows main
sequence donors of $1 .. 2 \, \msun$ to fill their Roche lobes still
on the MS (cf.\ Fig.~1 in Podsiadlowski et al.\ 2001).
A range of masses framed by the critical interval for TTMT will
therefore in many cases include the lowest mass where such a period is
still reached on the MS, i.e.\ lead to an extremely evolved donor with
hardly any hydrogen left in its core. In this simplified picture, we
thus obtain automatically two very different groups of CVs: Normal
one, showing a period gap, a bounce period, and a frontline at the
current period minimum, and in addition a very evolved looking group
derived from more massive secondary stars via a TTMT phase, many of
which have been reduced to the He-rich donors in AM CVn systems. 

We do not consider the assumption for such a post-CE configuration to
be very special. In fact, even at $P_{\rm post-CE} = 8 \, {\rm hr}$
system would not have been born below the gap in the age of the galaxy.
Given the current status of known pre-CVs, no system but MT Ser (whose
orbital period in still debated, cf.\ Bruch 2002) would indicate such a
short birth period after the CE.

It should however also be made clear that the whole issue of TTMT is
independent the way systems come into contact. The examples shown
allow for a mixture of the two groups, e.g.\ TTMT in systems driven by
magnetic braking. The important effect of TTMT is to allow more
massive, and possibly evolved systems to finally evolve towards
shorter orbital periods and thus to become CVs, even if no angular
momentum loss was operating (or strong enough) at their initial mass.


\section{Answers to various CV and LMXB problems}

Before addressing various problems and how the TTMT population in
particular can help to overcome these, we should take a careful look 
to understand what CVs we do actually observe.


\subsection{What do we actually see when looking at CV data}

Usually distributions of different subclasses and types of CVs are
compared over orbital period, so the issue of phase space density needs 
to be addressed. Low $\dot{M}$ -- or more accurately low $\dot{P}$ --
enhances the occurrence of certain objects. 
Thus it may seem plausible that above the period gap TTMT systems
(i.e.\ evolved donor stars) may dominate the whole population, whereas
below the gap (almost) normal looking CVs are the rule. 
Therefore such a bimodality does not automatically lead to a
contradiction with an observational agreement of the standard model. 

When it comes to the vicinity of the period gap, one possibly useful
way of overcoming the dreaded selection effects could be to
specifically analyse the subclass of AM Her magnetic systems. 
Finally we have to bear in mind that the CV distribution is not a
static one, but rather influenced by various temporal evolution
effects (generating the observed $P_{\rm min}$, modifying the current
masses of systems undergoing a TTMT, {\ldots}).


\subsection{Generic properties of TTMT tracks}


\subsubsection{White dwarf masses}

Let us assume for simplicity that all WDs in pre-CVs have initially
$\sim 0.6 \, M_{\sun}$. Those who pass through a TTMT can at least
during their super-soft phase grow by
$\sim \mbox{few} \, 0.1 \, M_{\sun}$. An example of a systems just
having left TTMT is  AE Aqr (Schenker et al.\ 2002), whose WD has a
claimed mass of $0.89 \, \msun$. Under special circumstances some may
even grow further and become SN Ia (cf.\ Langer 2002). 

We might therefore expect (or rather, predict) that post-TTMT CVs
would have larger $M_1$ on average. A larger number of weak cases of
TTMT (like the one shown in track 1) might actually smear out such an
effect, making it more difficult to establish such a relation (as
would a wide initial WD mass distribution). 

One of the immediate implications of this would be regarding nova
outbursts in the two groups: More massive WDs are supposed to ignite
earlier and more frequent, thus possibly leading to a preferential
detection of classical novae in post-TTMT systems. 

Similarly NSs may grow beyond their limiting mass and collapse into a
BH, or at least appear to be significantly more massive than a single
NS. Various cases (e.g.\ XTE J2123-0547) show hints of such an
overmassive primary.


\subsubsection{Chemical abundances, in particular the C/N ratio}

The surface abundances of the donors star (or equivalently the
accreted matter of disk or stream) can contain a clear signature of
CNO processed material. We interpret such observations as clear
indications of initially massive, evolved stars.

We have to caution however that there is also the possibility of
contamination by the WD (e.g.\ via novae) that needs to be excluded,
in particular when analysing the WD spectrum alone. 

A crucial point is, that only very evolved stars on the MS which
lose their envelopes rapidly before becoming fully convective can
reach extreme C/N ratios (similar to CNO equil.). In less extreme
cases some mixture with solar envelope material will take place,
leading to dilution and a value of the C/N ratio between solar and the
CNO equilibrium value. 
The required large $M_2$ (with a large mass fraction having approached
CNO equilibrium abundances of C and N) and a strong TTMT are both
natural consequences of the pre-CV scenario described above. This
picture is confirmed in AE Aqr (with the largest apparent C/N ratio
among CVs), where these requirements come also from its current
properties like mass, spin, etc.

Although not quite accurate, we feel confident to infer abundance
ratios from the line ratios of C{\sc iv} and N{\sc v} in the UV.
Mauche, Lee \& Kallman (1997) have analysed IUE data of a large number
of CVs. Besides some objects showing a very weak C line, many CVs
only have a slight increase in the C/N ratio, which might still carry
information about their nuclear past.

An important question (or challenge) to observers will therefore lie
in the answer to the question, how many CV 
show C/N anomalies similar to \mbox{AE Aqr} and \mbox{V1309 Ori}
(cf.\ King et al.\ 2001), and whether we will be able to reach a stage
of quantitative comparison between theory and observations. 


\subsection{A peak at the list of problems}

\begin{figure}
\plotfiddle{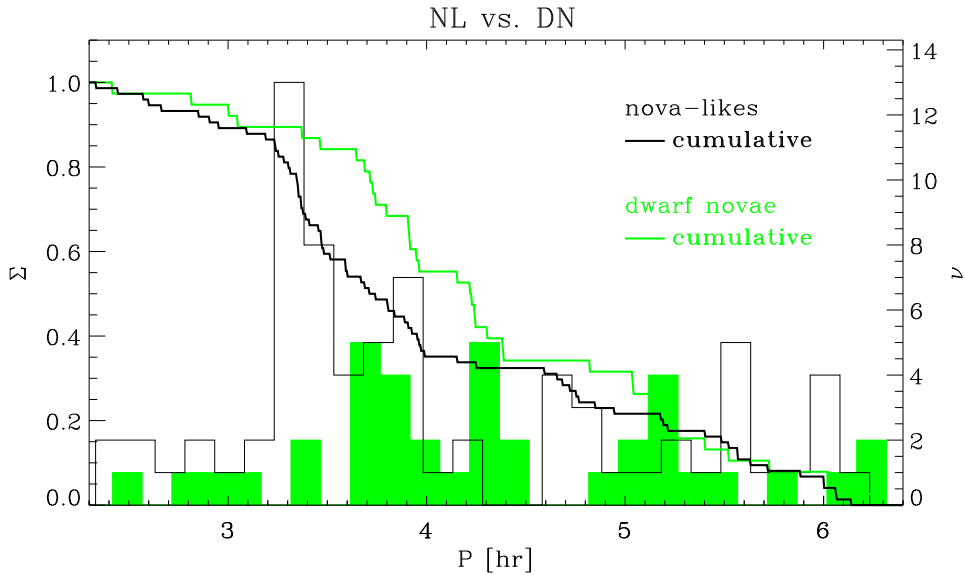}{7cm}{90}{55}{55}{220}{-40}
\caption{Distribution of dwarf novae (DN) and nova likes (NL) above
the period gap, with data taken from Ritter \& Kolb (1998). Histograms
show the number of systems per orbital period bin $\nu$, whereas the
lines represent the relative cumulative distribution with the scale on
the right vertical axis. Note the lack of DNs shortwards of 
$3.5 \, {\rm hr}$.}
\end{figure}

Many problems in the standard model are instantly solved assuming the
scenario sketched in this and the previous contribution. At the very
least, many of the issues look very different now, and can possibly be
overcome easily. 

To name but a few, several puzzles in connection with the observed
period minimum disappear when interpreted as an age effect (as
discussed in King \& Schenker 2002). Evolved donors naturally appear
above the period gap, and individual systems like AE Aqr or V485 Cen,
as well as a whole class of AM CVn binaries find their proper place
among the family of CVs and related objects. 

Let us now take a closer look at one particular, tricky example.

\subsubsection{The DN-NL problem just above the period gap}

\begin{figure}
\plotfiddle{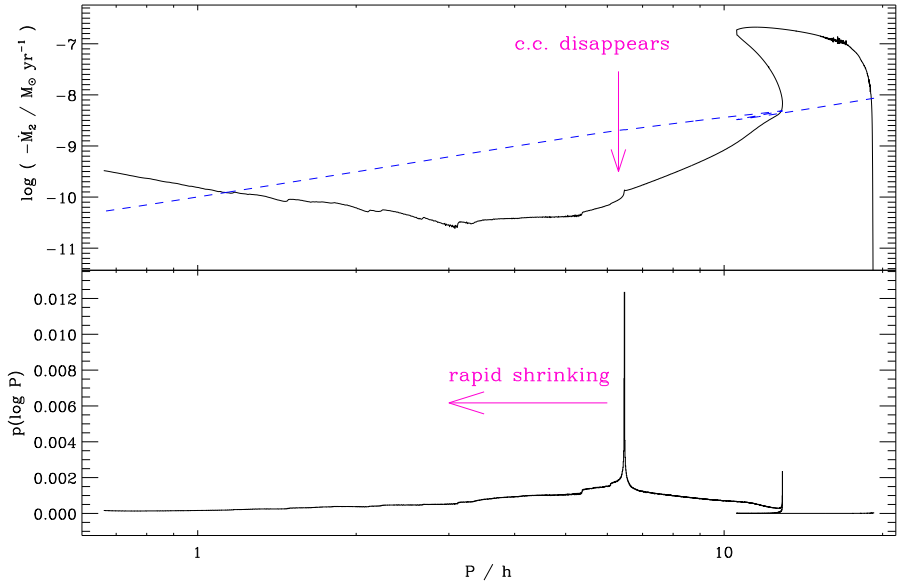}{7cm}{90}{55}{55}{220}{-40}
\caption{Evolution of track 2 in Fig.~1 shown together with the
systems intrinsic observational probability in the panel below. 
The additional dashed line marks the boundary for disk instability
(see text). The hump around the spike in the lower panel
(which is associated with the disappearance of the convective core of
the donor) reflects the enhanced chance of seeing a system at period
longer than $3 .. 4 \, {\rm hr}$.} 
\end{figure}

Figure 3 seems to suggest an apparent lack of dwarf novae (DNe) in the 
period range of $3 .. 3.5 \, {\rm hr}$, shown by comparing the 
distributions of DNe and nova-likes (NL) both binned and cumulative.
In a way this may also be described as a slightly different upper edge
of the period gap for DNe (as opposed to NL or the all CV subclasses
combined). This is very confusing mostly because in the region
immediately above the gap the critical mass transfer rate separating
stable from unstable disks is believed to be larger than the one
required to form the observed period gap in the interrupted magnetic
braking picture. In other words: theoretically we would expect to find
only DNe just above the gap, i.e.\ almost the opposite to what is
observed. 

This fairly long-standing problem still lacks any compelling
explanation (Shafter 1992).
Our contribution towards a possible solution of this paradoxon is
based on the phase space density argument introduced above.
In fact the observational probability of finding a system in a certain
period range in inversely proportional to
\begin{equation}
  \frac{\der \ln P}{\der t} \sim \frac{1}{2} 
    \left( 3 \, \zeta_2 - 1 \right) \frac{\der \ln M_2}{\der t}
  \, ,
\end{equation}
i.e.\ besides $\dot{M}_2$ it also depends heavily on the donor stars
mass--radius exponent $\zeta_2$.
A close look at track 2 in Fig.~4 together with its critical stability
line for the disk shows, that the system appears as a DN over almost
the entire CV period range between $1 .. 10 \, {\rm hr}$.
Thus this `extreme' TTMT provides a huge spike in observational
probability (bottom panel of Fig.~4) between $6 .. 7 \, {\rm hr}$ when
the convective core of the donor disappears and the orbital period
stalls (cf.\ Fig.~2). 
Shortly afterwards the star shrinks rapidly, rushing through period
space and thus is less likely to be observed as the DN which the system 
still would be. 
A suitable superposition of systems of this sort could provide enough
DNe to explain the set of low $\dot{M}_2$, long period systems that
`vanish' and apparently do not follow the paradigm required for the
gap formation at all. 

Obviously this has not solved the whole of the problem: `normal'
systems would still be expected to become DNe as they approach the
upper edge of the period gap. As these are the high $\dot{M}_2$
fraction of systems in our picture, the stability criterion would have
to be modified accordingly to stabilise all disks in these systems.
Clearly we still have a way to go before fully understanding this
phenomenon.

\subsubsection{Some of the not-so-well-hidden new problems}

A cautionary note should be made to some of the assumptions
made in arriving at the above solutions. 
The post-CE period is most likely not constant in pre-CVs. This may be
derived from observational data (Ritter \& Kolb 1998), indicating
longer periods for more massive systems, however sparse the actual
pre-CV data may be. This is also supported by theoretical models,
e.g.\ de Kool \& Ritter (1993). Various {\em ad hoc} slopes and shifts
could modify Fig.~4 significantly and lead to {\em very} different
proportions and properties of the two groups. 

Another possible point of concern is the clump of unusual looking
systems around $P_{\rm orb} \simeq 80 \, {\rm min}$, claimed to be
post-bounce period systems. Apart from the fact that (as discussed in
King \& Schenker 2002) this interpretations has a number of problems
as well, only a more accurate determination of masses, spectral types,
and composition of these systems can unravel the remaining mystery
about them. Being at the frontline of the `normal' population group
may imply some special properties (e.g.\ with respect to metallicity)
which lead to some unexpected features.

Most seriously, however, we do not understand magnetic braking, even
more so after the presentation given at this meeting. Thus we are
effectively left without a properly working model for the period gap.
Clearly the approach taken herein, i.e.\ to simply assume a
sufficiently strong angular momentum loss mechanism of unknown origin
and (nowadays) relatively unmotivated specification (Verbunt \& Zwaan)
is no longer satisfactory.


\section{Conclusion}

The concept of allowing and even integrating thermal timescale mass
transfer into the general evolution of CVs has proven very successful. 
It allows us to explain many individually {\em strange systems} (AE
Aqr, V1309 Ori, {\ldots}, \mbox{AM CVn}!?) that would otherwise break
the coherent picture of the standard model. 

Closely related is the realization of a second formation channel for
CVs, namely forming the CV by nuclear evolution followed by the fast 
TTMT, that allows angular momentum loss to take over driving the
evolution only after this initial phase.
The bimodality of CVs formed in both ways thus naturally provides an 
evolved population and a spread in $\dot{M}_2$ above the period gap
{\em without necessarily destroying the gap and the population below
the gap}. 

We consider it an important strength of this scenario, that a large
number of things appear to fall into place now. In particular, the
period minimum can be understood as an age effect, CVs below the gap
are generally not born there (but at rather higher masses), and 
ultra-short period binaries (AM CVn for WD primaries, as well as the
corresponding LMXBs) derive from fairly evolved \& massive donor stars 
after passing through a CV-like phase. 
The overall differences between CVs and LMXBs appear diminished and
due to well understood differences between the two primary types. 
It is reassuring that CVs and LMXBs do follow the same fundamental
evolutionary principles after all.



\acknowledgements{Theoretical astrophysics research at Leicester is
supported by a PPARC rolling grant.}

\end{document}